\documentclass[12pt]{article}

\usepackage{epsfig}

\usepackage{latexsym}

\begin{document}

\begin{titlepage}

\baselineskip 24pt

\begin{center}

{\Large {\bf Suggestions for Identifying the Primary of Post-GZK Air Showers}}

\vspace{.5cm}

\baselineskip 14pt

{\large Jos\'e BORDES}\\
jose.m.bordes\,@\,uv.es\\
{\it Departament Fisica Teorica, Universitat de Valencia,\\
  calle Dr. Moliner 50, E-46100 Burjassot (Valencia), Spain}\\
\vspace{.2cm}
{\large CHAN Hong-Mo}\\
chanhm\,@\,v2.rl.ac.uk \\
{\it Rutherford Appleton Laboratory,\\
  Chilton, Didcot, Oxon, OX11 0QX, United Kingdom}\\
\vspace{.2cm}
{\large TSOU Sheung Tsun}\\
tsou\,@\,maths.ox.ac.uk\\
{\it Mathematical Institute, University of Oxford,\\
  24-29 St. Giles', Oxford, OX1 3LB, United Kingdom}

\end{center}

\vspace{.3cm}

\begin{abstract}

A procedure is suggested for systematically narrowing the choice of possible
primaries for (UHECR) air showers with energies beyond the 
Greisen-Zatsepin-Kuz'min cut-off of $4 \times 10^{19}$ eV.

\end{abstract}

\end{titlepage}

\clearpage

\baselineskip 14pt

\setcounter{equation}{0}

The continued observation in the last few years of ever more air showers 
\cite{Agasa,Havepark,Zas} with energies beyond the Greisen-Zatsepin-Kuz'min 
(GZK) cutoff \cite{Greisemin} has further increased the credibility of 
these events and the urgency for answering the question of their likely 
origin.  Since protons and other atomic nuclei, the primaries of lower 
energy air showers, are thought not to survive journeys longer than 50 
Mpc through the cosmic microwave background with energies greater than 
$4 \times 10^{19}$ eV, it follows that the primary of any shower observed 
with energy above that cutoff has either to originate within a 50 Mpc 
radius or else to be some particle other than a proton or an atomic nucleus.  
In either case, it signifies new physics since, as of today, we know neither 
of any likely astrophysical sources capable of emitting such energetic 
particles within a distance of 50 Mpc nor of any particle capable of 
surviving a long journey through the 2.7 K microwave background and yet 
produce such an energetic air shower on collision with an air nucleus.

Obviously, one ingredient of first importance for resolving the mystery of
post-GZK air showers is the identity of the primary particle producing
them.  The purpose of this paper is to suggest a possible procedure, from
the theoretical point of view at least, for systematically narrowing the 
choice of possible candidates with the aim of eventually zeroing in on 
the actual identity of the post-GZK primary.  We believe that this will go 
a long way towards explaining the effect and open a window into a new area 
of physics which may as yet be entirely unconceived.   

A first question to ask is whether the primary particle responsible for
post-GZK air showers is or is not the proton or an atomic nucleus.  If it
is, then the explanation for post-GZK air showers would seem to lie, by
the reasoning above, in finding some hitherto unsuspected nearby source
either in the form of novel celestial bodies capable of producing cosmic 
rays of such high energies, or else, as also suggested, from the decay 
of ultra-massive particles \cite{Sarkeretal} or from the annihilation of 
relic neutrinos \cite{Gelmini} left over from the Big Bang.  At our present 
level of understanding, this seems the only conclusion we could draw unless 
we are willing to consider more drastic modifications of the basic laws 
of physics, such as the intriguing suggestion in \cite{Coleshow,Gonzalez} 
of a violation of Lorentz invariance at ultra high energies, so that the 
GZK-cutoff is itself invalidated.  On the other hand, if the primary 
particle of post-GZK air showers is not the proton or an atomic nucleus, 
but some other particle coming from a distant source, then the explanation 
would seem to lie in the realm of new particle physics beyond the current 
Standard Model, either in terms of a new interaction of some known particle 
(such as a new strong interaction for neutrinos at high energy as we ourselves
advocate \cite{airsho1,airsho2,airsho3} (see also \cite{Domokosetal}), or 
else in terms of some altogether new particle not yet experimentally observed 
(such as light gluinos etc.\ \cite{Farraretal}).\footnote{There has been a 
claim in the literature \cite{Halzenetal} that particle physics explanations 
for post-GZK showers in general do not work.  We believe this claim to be 
ill-founded for reasons given in \cite{fcnc,airsho3}.}

The first question we pose ourselves, therefore, is whether one can devise 
some means for distinguishing the proton or an atomic nucleus from some 
other particle as the primary of post-GZK air showers.  (We shall assume
for the sake of economy that, if not the proton or an atomic nucleus, then 
it is only one other type of particles acting as primaries for post-GZK 
air showers, which is reasonable given that it is already hard enough 
to find just any particle to suit the purpose.)  Our proposal is the
following, the idea for which has already been alluded to in an earlier 
paper \cite{airsho2}.  The penetrating power of a particle through the
atmosphere is determined by its cross section with air nuclei, which cross
section is itself a characteristic of the particle.  Now, air showers at 
energies shortly below the GZK cutoff are thought to be mostly protons,
which has a cross section with air nuclei equivalent to a flux attenuation 
constant of around $60 \rm{gm}/\rm{cm}^2$ implying an average penetration 
depth of about 25 km.  If air showers above the GZK cutoff are also produced
by protons as primaries, then the penetration depth should remain similar.  
On the other hand, if the primaries for post-GZK air showers are some 
different particles, which have no reason to have the same cross section 
as protons with air nuclei, then the average penetration depth would show 
a near discontinuous change as the energy crosses the GZK cutoff.  

\begin{figure}
\centering
\includegraphics[angle=-90,scale=0.55]{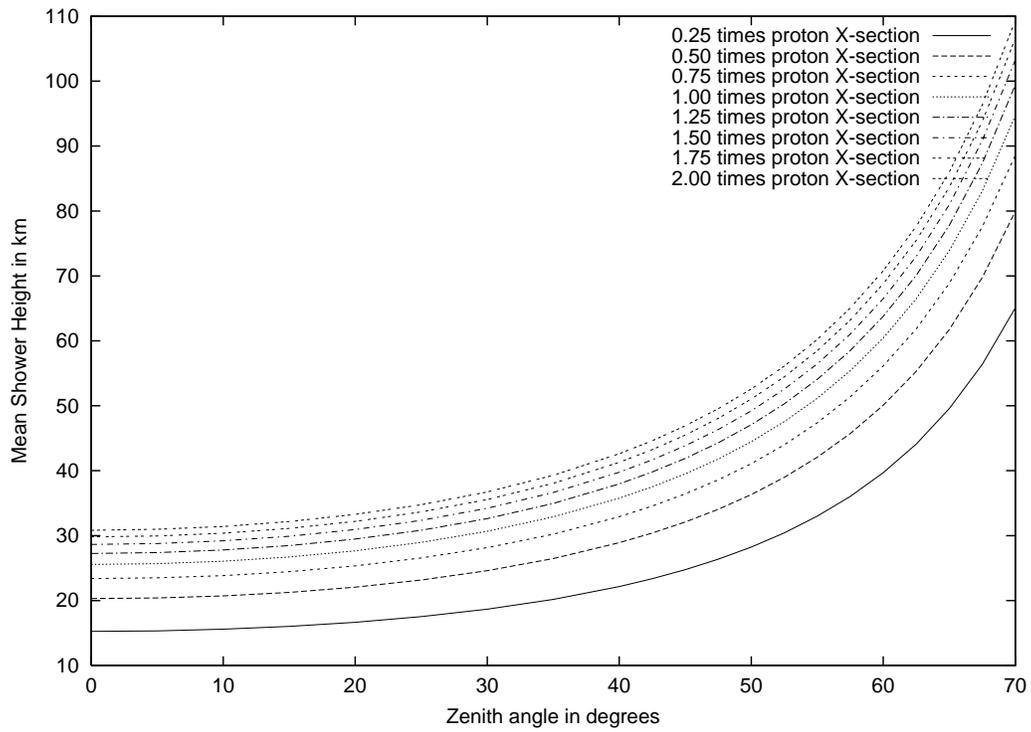}
\caption{Average penetration depth $\bar{r}_1$ as a function of the incident
zenith angle $\theta$ for varying cross sections of the primary.}
\label{r1bartheta}
\end{figure}

Explicitly, the probability for a primary particle to penetrate to a depth
$r$ at an angle $\theta$ to the zenith and effecting a collision there to
produce an air shower is easily shown to be \cite{airsho2}:
\begin{equation}
F(r,\theta) = K(\sigma) \rho(h(r,\theta)) f(r,\theta),
\label{Frtheta}
\end{equation}
where $K(\sigma)$ is the flux attenuation constant:
\begin{equation}
K(\sigma) = (N/A) \sigma,
\label{Ksigma}
\end{equation}
with $N$ being the Avogadro number, $A$ the atomic number of the air nucleus,
and $\sigma$ the incident particle-air nucleus cross section. 
The function $f(r,\theta)$
is the attenuated flux at the point $(r,\theta)$:
\begin{equation}
f(r,\theta) = f_{inc} \exp \left\{ K(\sigma) \int_{\infty}^r dr' 
   \rho(h(r',\theta)) \right\},
\label{frtheta}
\end{equation}
and $\rho(h)$ is the air density at height $h$ in cm above sea-level, which
can be roughly parametrized as \cite{databook}:
\begin{equation}
\rho(h) = 1.2 \times \exp (-h/h_0) \times 10^{-3}\ \rm{gm}/\rm{cm}^3,
\label{rhoh}
\end{equation}
where:
\begin{equation}
h_0 \sim 7.5 \times 10^5\ {\rm cm},
\label{hzero}
\end{equation}
and $h$ is given in terms of $r,\theta$ and the radius of the earth $R$ as:
\begin{equation}
h(r,\theta) = \sqrt{R^2 + 2rR \cos \theta + r^2} - R.
\label{hrtheta}
\end{equation}
With these formulae, it is easy to evaluate the average penetration depth
at any incident angle $\theta$ for any given value of the primary particle
cross section $\sigma$ with air nuclei.  In Figure \ref{r1bartheta}, this
average penetration depth $\bar{r}_1(\theta)$ so calculated is plotted as 
a function of $\theta$ for various cross sections $\sigma$ of the primary 
particle with air nuclei.  One sees that $\bar{r}_1$ is quite sensitive to
$\sigma$.  For example, for $\theta = 0$, one obtains for the benchmark value
of $K = 60\ \rm{gm}/\rm{cm}^2$ appropriate for protons, a value for $\bar{r}_1$
of 25.1 km, but for a particle with half that cross section, an $\bar{r}_1$
of only 19.8 km, a drop by about 6 km.  Hence, if the primary for post-GZK
showers is a particle with only half the cross section of the proton, our
arguments above suggest that $\bar{r}_1(\theta)$ for any $\theta$ as a
function of the primary energy would show a sudden drop of that order as
it crosses the GZK threshold.

However, given the scarcity of post-GZK events, it looks likely that for
many years to come, one will not have collected enough events to evaluate
$\bar{r}_1$ as a function of energy for a fixed value of the zenith angle 
$\theta$.  Fortunately, the dependence of $\bar{r}_1$ on $\theta$ is rather
trivial and can be readily integrated out, as follows.  Define for any
shower with primary vertex (i.e.\ point of first interaction) at $r,\theta$
an equivalent vertical height $h_1$ as follows:
\begin{equation}
\int_\infty^{h_1} dh \rho(h) = \int_\infty^r dr' \rho(h(r',\theta)),
\label{h1def}
\end{equation}
namely the height at which a vertical primary would have traversed the
same amount of air as the actual particle incident at the angle $\theta$
has traversed by the time it reaches the point $r,\theta$.  By definition,
we have of course:
\begin{equation}
\frac{dr}{dh_1} = \frac{\rho(h_1)}{\rho(h(r,\theta))},
\label{drdh1}
\end{equation}
so that if we define an average equivalent vertical height $\bar{h}_{eq}$
as:
\begin{equation}
\bar{h}_{eq} = \int_0^\infty dr' h_1(r',\theta) F(r',\theta)/
   \int_0^\infty dr' F(r',\theta),
\label{heqbardef}
\end{equation}
we have easily:
\begin{equation}
\bar{h}_{eq} = \int_{r'=0}^\infty dh'_{eq} h'_{eq} F(h'_{eq},0)/
   \int_{r'=0}^\infty dh'_{eq} F(h'_{eq},0).
\label{heqbar}
\end{equation}
Compare this with:
\begin{equation}
\bar{h} = \int_{h'=0}^\infty dh'  h' F(h',0)/ \int_{h'=0}^\infty dh' F(h',0),
\label{hbar}
\end{equation}
namely the average penetration depth of air showers coming down vertically.
One sees that the two quantities differ only by the lower limits of the
integrals, i.e. $h' = 0$ for the latter but $h'$ at $r' = 0$ for the former.
However, given that the weight function $F(h,0)$ is already very small by
the time it reaches ground level, one can to a good approximation neglect
this difference in the lower limits and put:
\begin{equation}
\bar{h}_{eq}(\theta) = \bar{h}.
\label{h1bara}
\end{equation}
In other words, the average equivalent vertical height for any incident
angle is to a good approximation the same as the average height of air
showers coming down vertically and independent of the incident angle.
Thus, instead of evaluating the average penetration depth for each value
of $\theta$, we can evaluate the equivalent vertical height $h_1$ averaged 
over all events of whatever incident angle $\theta$, and this will have 
the same near disconituity at the GZK cut-off energy as for the average 
height of showers coming down vertically.  As far as statistics
is concerned, therefore, the present amount of data could already be
enough for uncovering the phenomenon if post-GZK primaries are indeed
particles other than the proton with a different cross section.

\begin{figure}[p]
\centering
\includegraphics[angle=-90,scale=0.55]{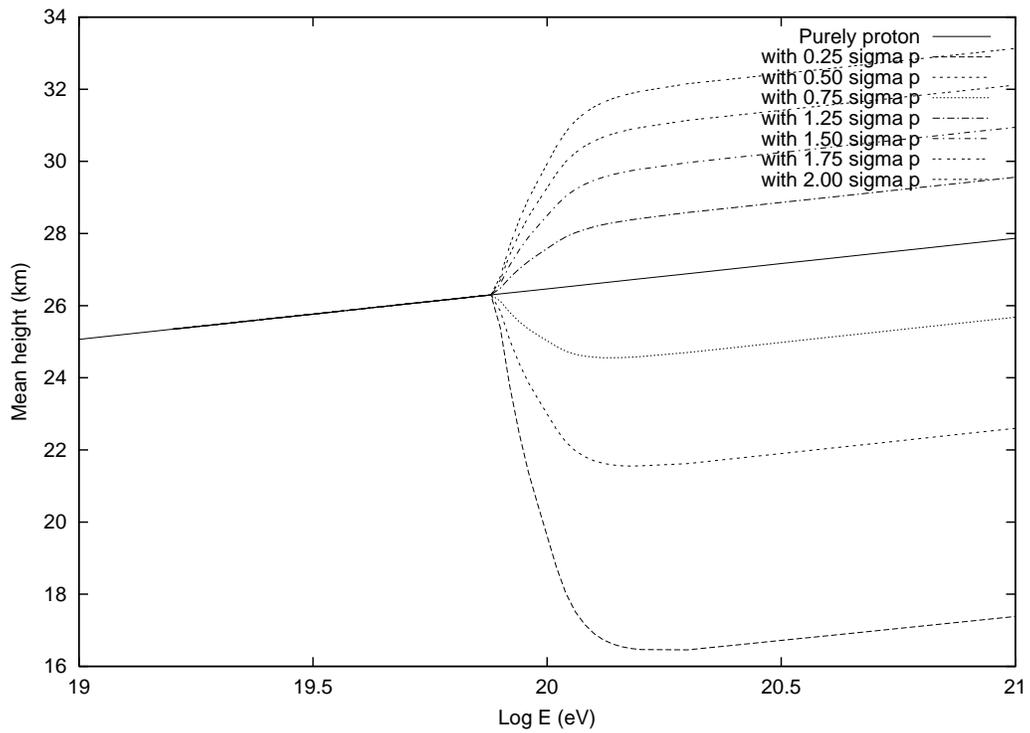}
\vspace*{5mm}
\caption{Average equivalent vertical height of air showers as a function
of the primary energy across the GZK cut-off assuming post-GZK primaries
of varying cross sections.}
\label{Muigraph}
\end{figure}

Of course, the GZK cut-off not being abrupt, and particle cross sections
themselves being energy dependent, albeit only slowly, the expected near 
discontinuity in $\bar{h}_{eq}$ may possibly be masked and not be readily
observable.  To test this, we have calulated $\bar{h}_{eq}$ using the 
above formalism but under the following more realistic assumptions: (i)
that the total cosmic ray spectrum, say $N$, is given by the fit of Bird 
et al. \cite{Birdetal}:
\begin{equation}
\frac{dJ}{dE} \sim (E/E_a)^{{-\gamma}_1} + (E/E_a)^{{-\gamma}_2},
\label{Birdfit}
\end{equation}
with the recent best fit parameters \cite{Takedaetal}:
\begin{equation}
\gamma_1 = 3.16, \ \ \gamma_2 = 2.78, \ \ E_a = 10^{19.01}\ {\rm eV},
\label{Takedafir}
\end{equation}
(ii) that the proton spectrum $N_p$ is cut-off by the GZK effect according 
to the calculation reported in \cite{Takedaetal}, (iii) that the difference 
$N_X = N - N_p$ is made up by a particle $X$ having a cross section $\sigma$ 
with air nuclei different from that of the proton, and (iv) that the cross 
sections $\sigma_p$ and $\sigma_X$ both rise slowly with energy as say 
$\sim s^{0.08}$ as suggested by the fit of \cite{Donshoff}.  Then if 
${\bar h}_p$ is the average height of proton showers, and ${\bar h}_X$
that of $X$ showers, the average height ${\bar h}$ of all showers observed
would be 
\begin{equation}
{\bar h} = \frac{{\bar h}_p N_p + {\bar h}_X N_X}{N}.
\label{hbarbar}
\end{equation}
The result is shown in Figure \ref{Muigraph}.  One sees that in spite of
some smoothing out, the near discontinuity remains markedly noticeable.

Since the test involves taking averages of penetration depths, it cannot 
of course distinguish primaries of air showers on an event by event basis.  
However, if experiments can locate the primary vertex of an air shower 
with some accuracy, the test seems sufficient to provide the answer to
the question whether the primary for post-GZK air showers is or is not 
the proton.  The crucial question then is whether an experiment can be
devised to locate the primary vertex of an air shower with the required
accuracy.  This is an experimental question that we as theoreticians
cannot presume to answer.  Naively, however, it seems to us that among 
current designs it is the Fly's Eye experiment \cite{Birdetal}, or its 
derivatives such as HiRes \cite{HiRes} and AUGER \cite{Auger}, capable of 
mapping the development profile of an air shower which will have the best 
chance.  Although a shower will not presumably produce an observable 
light signal until some time after the first interaction, nevertheless, 
by studying the development profile of the shower in its initial stages 
with a simulation program which lays particular emphasis on reproducing 
these aspects, one should be able, we think, to locate the primary vertex 
with some confidence, hopefully to the accuracy of about a kilometer for 
our purpose.  

Indeed, if our aim is not as yet to measure the actual cross section of 
the post-GZK primary, but only to ascertain whether it is or is not a 
proton, then what is really essential is not the actual location of the 
primary vertex but only whether there is or is not a near discontinuity
in the energy dependence of $\bar{h}_1$.  In that case, any measurable 
property of air showers which depends sensitively on the height $h_1$ of 
first interactions could in principle be used instead of $h_1$ for the 
analysis, since it should also show a near discontinuity around the GZK 
cut-off so long as its functional dependence of $h_1$ is smooth.  Possible
examples are the equivalent vertical height at which light is first 
detected from the shower in a Fly's Eye type detector \cite{Birdetal},
or else the height of showers as estimated from extrapolating muon tracks
backwards in arrangements such as GRAND \cite{Grand}.  Whether such
alternatives are at all feasible can be decided only by simulations of 
shower development folding in the experimental details, on which question
we are incompetent to judge at this stage.

If the result of an analysis with $\bar{h}_1$ or any of its alternatives 
show a clear change in value at around the GZK threshold, then it can be 
taken as indication that post-GZK air showers are generated by some particle 
other than the proton.  If the measurement is accurate enough, particularly 
in the case of $\bar{h}_1$, it would be possible even to estimate the value 
of the cross section of this new primary with air nuclei and use it as one 
ingredient for identifying the particle.  On the other hand, if $\bar{h}_1$
remains perfectly smooth across the GZK threshold, showing no tendency at 
all for any sudden change, it can mean either it is still the proton which 
is reponsible for post-GZK air showers or that these showers are generated 
by some particle with cross sections very similar to the proton, which is 
possible although it would be somewhat of a coincidence.  In the latter 
case, therefore, one would be inclined to look for near galactic sources 
capable of producing protons of these ultra-high energies.  

Suppose however that the result of the test is positive so that a new
primary is indicated.  The natural inference then would be that these
particles come from distant sources for otherwise, if there are nearby 
sources capable of producing these new primaries with such high energy,
there is no reason they cannot produce as well protons of such energies
which will then give many more post-GZK proton showers in contradiction to 
the starting supposition.  That being the case, the new primary must have 
presumably zero electric charge or otherwise it would have interacted with
the 2.7 K microwave background and be degraded in energy.  Furthermore,
it was pointed out in \cite{Zas} that out of the 60 or so post-GZK
events with energy greater than $4 \times 10^{19} {\rm eV}$ detected by
AGASA \cite{Agasa}, there are 4 doublets and even 1 triplet of directional
coincidences, with the members of each multiplet differing in incident 
direction by less than 2.5 degrees.  The probability of these coincidences
happening by chance was estimated to be less than 0.3 percent, which 
suggests that members of the same multiplet originate from the same source.  
If this effect is genuine, then it again favours zero charge for the 
new primaries, for otherwise the members of a multiplet having different 
energies would have been deflected differently by the intergalactic 
magnetic fields and arrive on earth in quite different directions.  

One is thus reduced to seeking a stable neutral particle which can survive
a long journey through the microwave background but has yet a large enough
cross section with air nuclei as to be capable of producing a shower in the
earth's atmosphere.  As far as our present knowledge goes, of course, there 
is no such particle, which is the original source of the mystery.  Faced 
with this dilemma, one can consider two possibilities: either that the 
post-GZK primary is an already known particle having acquired some new 
interaction at these ultra-high energies, or that it is an altogether new 
particle which has never been experimentally identified before.  (Of 
course, it can also be that the post-GZK primary is both a new particle 
and one which has an unusual interaction only at ultra-high energies, but 
this possibility we shall tentatively discard as being uneconomical.)  For
the first alternative, there is only one candidate, namely the neutrino,
which is neutral, stable, and has no difficulty traversing the microwave
background over long distances with its energy intact. \footnote{Note that
a $10^{20}$ eV neutrino colliding with a photon in the microwave background
has a cm energy of only a few hundred MeV, far below the cm energy of a 
few hundred TeV for its collision with a nucleon in an air nucleus at 
which the new strong interaction is supposed to operate.}  For the other 
alternative, the field is of course wide open.    

The relative likelihood of these two alternatives can be contrasted if one 
accepts the directionally coincident multiplets of AGASA \cite{Agasa} cited 
in the preceding paragraph as genuine and as meaning that each multiplet 
originate from the same source.  Presumably, a source capable of generating 
particles of such energies as $10^{20}$ eV should also generate many more 
particles of lower energies, unless the ultra-high energy particles are 
produced at source by a mechanisms which operate only at these energies.  
Hence, if the post-GZK primary is a new particle, with by assumption no 
unusual interaction operative only at ultra-high energy, then coming from 
the directions defined by the multiplets of \cite{Agasa}, one expects many 
more showers at lower (pre-GZK) energies.  Whereas, if the post-GZK primary 
is a particle (old or new) which acquires its strong interactions with 
nuclear matter only at ultra-high energies of order $10^{20}$eV, then no 
enhanced collimation of pre-GZK showers in the directions of the AGASA 
coincident multiplets need be expected, since in the first place, these 
primaries of lower energies would not be copiously produced by the source, 
and even if they were, they would not, on arrival one earth, be strongly 
interacting enough with air nuclei to produce air showers.  So far, to 
our knowledge, no unusual collimation of pre-GZK energy showers has been 
reported by experiment along the directions of the coincident multiplets 
observed by AGASA.  We suggest that a search for the collimation effect 
be made.  If the AGASA coincident multiplets are real, then the effect 
could be decisive between the two alternatives considered; collimation 
favours a new particle as post-GZK primary, while no collimation favours 
the neutrino alternative.

To proceed further, we think we shall have to rely more on particle physics 
than air shower physics in attempting to identify the post-GZK primary.
The reason is that, once a shower is produced, its later development will
depend only subtly on the primary interaction, and such subtleties are 
probably beyond the predictive power of present theories either of new 
particles or of new interactions.  Each theory, however, which claims to 
offer an explanation for post-GZK air showers should in principle be able 
to make concrete predictions in particle physics tied to the post-GZK
phenomenon, and these predictions can be subjected to tests by experiment.
As example, rather than explore the wide open field of new particles as the 
post-GZK primary, we shall concentrate now on the neutrino alternative. 

For the neutrino to act as post-GZK primary, it has to acquire at ultra-high 
energy a strong interaction.  The simplest scenario in which this can 
happen is when neutrinos interact via the exchange of some heavy particle 
\cite{airsho1,airsho2,airsho3}.  Then at energies much below the mass of the
exchanged particle, the effect of its exchange will be suppressed, just as 
``weak'' interaction used to be considered weak at energies much below the $W$
mass.  However, when energies becomes comparable to the mass of the exchanged
particle, this suppression will no longer operate and the interaction
strength will then be characterized just by the coupling constant, which 
can in principle be strong.  Such a scenario arises naturally in schemes 
where generation is supposed to originate from a broken `horizontal'' gauge 
symmetry, since any particle carrying the generation index, including in 
particular the neutrino, will then interact via the exchange of the gauge 
bosons associated with the horizontal symmetry.  These bosons must have 
large masses or otherwise they will give flavour-changing neutral current 
(FCNC) effects violating existing experimental bounds.  On the other hand,
they cannot be too heavy if their exchange is to be responsible for the 
interaction giving rise post-GZK showers.  Indeed, a rough upper bound
on the mass of these bosons would be given by the cm energy of a $10^{20}$
eV primary impinging on a stationary air nucleus or a nucleon within it,
which is of order 500 TeV.  Now an upper bound on the gauge boson mass 
corresponds in the present scenario to lower bounds on FCNC effects such 
as rare decays (e.g. $K_L \rightarrow e^\pm \mu^\mp$) or $\mu-e$ conversions
(e.g. $\mu^-\;Ti \rightarrow e^-\;Ti$) measured at low energy, on
which effects 
there are already very stringent empirical limits, which has thus to be 
satisfied for consistency.  For the particular scheme that we ourselves
suggested \cite{airsho1,airsho2,airsho3}, the predictions on FCNC effects
along these lines are detailed and explicit, and they have been tested and
found to survive all existing experimental bounds in all the cases studied 
\cite{fcnc,mueconv,condugen}.  One concludes then that the scheme remains
for the moment a viable explanation for post-GZK air showers.  But the
predictions are in some cases rather close to the existing empirical limits 
so that experiment in the near future should reveal whether the scheme's
present viability will be maintained.

Demands could be made on other schemes for similar concrete and testable 
predictions in particle physics, which are tied to their proposed explanation 
of post-GZK air showers.  The quality of these predictions and their ability 
to survive experimental tests could then be used as criterion to judge the 
relative merit and credibility of the schemes, whether with new particles 
or with new interactions for neutrinos.

For the latter category assuming new interactions for neutrinos, there is 
still one more important theoretical criterion that any scheme will
have to satisfy.
The exchange of heavy gauge bosons is not by any means the only scenario
that one can imagine for giving a new strong interaction to neutrinos at 
ultra-high energy.  For instance, \cite{Jainetal} assumes that space-time 
has hidden dimensions which give rise to heavy spin 2 bosons coupling to 
neutrino giving them thus strong interactions for neutrinos at ultra-high
energy.  However, as emphasized in \cite{airsho2}, for the neutrino to 
have a hadron-sized cross section with air nuclei so as to produce air 
showers a strong interaction by itself is not enough. It has to interact 
not just strongly but also coherently with the nucleus (or at least with 
the nucleons inside it).  Interactions due to heavy particle exchange are 
short-ranged, and if the neutrino interact strongly but only incoherently at 
short range with the partons inside the nucleon, the nucleon will appear to 
the neutrino just as a collection of black dots representing the partons, 
each dot with a size typified by the interaction range, which will be far 
from enough to give the interaction a hadron-sized cross section.  Thus, 
one further necessary criterion for judging the merit of proposed schemes 
with neutrinos as the primary of post-GZK air showers is whether they can 
explain a hadron-sized cross section for neutrinos.  The particular scheme 
we suggested \cite{airsho2,airsho3} based on what we call the Dualized 
Standard Model (DSM) \cite{dualgen} contains just such a mechanism, albeit 
poorly understood as yet, for obtaining for neutrinos a hadron-sized cross 
section sufficient for producing air showers.  

Much need yet be done both experimentally and theoretically before one
can expect to unravel the mystery of post-GZK air showers, but it appears
from the above discussion that, with enough patience, it may not be such 
an impossible task.

\end{document}